\documentclass[superscriptaddress,amsmath,amssymb,aps,prb,twocolumn]{revtex4-2}
\usepackage{comment}
\usepackage{graphicx}
\usepackage{dcolumn}
\usepackage{bm}
\usepackage{amssymb}
\usepackage{amsmath}
\usepackage{sublabel}
\usepackage[utf8]{inputenc}
\usepackage{hyperref}
\usepackage{braket}
\usepackage[usenames, dvipsnames]{color}
\usepackage[english]{babel}
\usepackage[T1]{fontenc}
\usepackage{bbm}

\usepackage{float}
\usepackage{verbatim}
\hypersetup{colorlinks = true, linkcolor=blue, citecolor=blue, urlcolor=blue}
\usepackage{xcolor}

\begin{document}
\preprint{APS/123-QED}
\title{Symmetry-driven spin anisotropic magnetotransport in quantum spin Hall insulator $WTe_2-1T'$}
\author{Shrushti Tapar}
\affiliation{Department of Electrical Engineering, Indian Institute of Technology Bombay, Powai, Mumbai-400076, India}
\author{Bent Weber}
\affiliation{Division of Physics and Applied Physics, School of Physical and Mathematical Sciences,
Nanyang Technological University, Singapore 637371, Singapore}
\author{Saroj P. Dash}
\affiliation{Department of Microtechnology and Nanoscience, Chalmers University of Technology, SE-41296,Goteborg, Sweden}
\author{Shantanu Mukherjee}
\affiliation{Department of Physics
Indian Institute of Technology Madras
Chennai, Tamil Nadu 600036, India}
\affiliation{Center for Atomistic Modelling and Materials Design, IIT Madras, Chennai 600036, India}
\affiliation{Quantum Centers in Diamond and Emergent Materials (QCenDiem)-Group, IIT Madras, Chennai, 600036 India}
\author{Bhaskaran Muralidharan}%
 \email{bm@ee.iitb.ac.in}
\affiliation{Department of Electrical Engineering, Indian Institute of Technology Bombay, Powai, Mumbai-400076, India}

\begin{abstract}
We present a comprehensive magnetotransport analysis of monolayer 1T$^\prime$-WTe\textsubscript{2}, highlighting the role of nonsymmorphic symmetries in governing edge-state spin behavior. By comparing the electronic transmission in nanoribbons with edges along the crystallographic $\hat{y}$ and $\hat{x}$ directions, our analysis reveals a pronounced anisotropy in the magnetic field response. The $\hat{y}$-edge ribbon exhibits significant spin splitting of edge-state bands in both energy and momentum space, along with a strong angular dependence of the conductance. The observed magnetotransport response indicates a spin quantization axis that aligns with the out-of-plane spin quantization axis reported in previous experimental studies. In contrast, the $\hat{x}$-edge ribbon shows negligible spin splitting under magnetic fields, which is attributed to nonsymmorphic symmetries such as glide mirror and screw rotation, that protects degeneracies along the $\Gamma$–$X$ direction, even when time-reversal symmetry is broken. The energy-resolved current density and angular transmission analyses confirm that this anisotropy originates from edge states, while bulk states remain largely insensitive to the field orientation. Our results establish direct transport-spectroscopy based evidence of nonsymmorphic-symmetry-protected spin degeneracy in the 1T$^\prime$-WTe\textsubscript{2}, and underscores its promise for spintronic devices that leverage symmetry-protected and directionally selective transport channels.
\end{abstract}

\maketitle
\section{\label{sec:level1}Introduction}

\indent Tungsten ditelluride (WTe$_2$) monolayers in the 1T$'$ phase have garnered significant attention because of their unique electronic and spintronic applications. They adopt an orthorhombic, centrosymmetric crystal structure with low in-plane symmetry. Strong intrinsic spin-orbit coupling (SOC) drives a non-trivial band inversion~\cite{muechler2016topological,choe2016understanding,jia2017direct} and supports robust quantum-spin-Hall edge states~\cite{qian2014quantum,fei2017edge,tang2017quantum,shi2019imaging,wu2018observation,lodge2021atomically,jia2022tuning}. Beyond the topological properties, 1T$'$-WTe$_2$ exhibits a rich catalog of tunable phenomena, including gate-induced superconductivity~\cite{symmetry_transport_exp,sajadi2018gate,xie2020spin,song2024unconventional}, gate- or strain-driven phase transitions~\cite{que2024gate,ji2023influence,zhao2020strain}, electrostatic band-gap modulation~\cite{maximenko2022nanoscale}, and pronounced anisotropic magnetoresistance (AMR)\cite{safeer2019sot,macneill2017symmetry,macneill2019angular}. Additionally, current-induced magnetization in inversion-broken structures~\cite{wang2019current,zhang2018electrically} positions 1T$'$-WTe$_2$ as a multifaceted platform for spin-orbit-torque magnetic random-access (SOT-MRAM) memories \cite{zhao2020observation,liu2023crystallographically}, and spin-valve architectures.

\indent Spin-dependent transport in 1T$'$-WTe$_2$ results from a nontrivial interplay between crystal symmetry and the orbital origin of SOC, which together influence spin texture, spin splitting, and magnetic anisotropy. Experiments have reported an out-of-plane, $\hat{y}$–$\hat{z}$ plane-canted spin axis~\cite{spin_texture1,spin_texture2}, while some theoretical studies suggest nearly momentum-independent SOC~\cite{shi2019symmetry}. These findings are supported by theoretical predictions of spin canting~\cite{garcia2020canted,arora2020cooperative} and tunable spin textures under external fields~\cite{shi2019symmetry}. However, a comprehensive symmetry-based framework that explains spin anisotropy along different lattice directions remains incomplete.

\indent Competing microscopic interpretations exist for the band inversion mechanism. Some studies attribute it to hybridization between the W $d_{xz}/d_{z^2}$ and Te $p_{x,y}$ orbitals~\cite{qian2014quantum,liu2014prl}, while others propose an intra-atomic inversion within the W $d$ orbitals~\cite{liu2018prx}. Among the resulting tight-binding models, a minimal four-orbital Hamiltonian constructed from W $d_{x^2-y^2}$ and Te $p_x$ orbitals~\cite{lau2019influence} has been extensively validated against density functional theory (DFT) and angle-resolved photoemission spectroscopy (ARPES) data. It reproduces the experimentally observed low-energy band structure and serves as a reliable framework for studying spin-resolved phenomena. Building on this picture, the present work theoretically investigates the influence of crystallographic symmetries, particularly nonsymmorphic operations, on the occurrence or suppression of spin splitting along different transport directions. The resulting anisotropy in the response to Zeeman fields determines the directional sensitivity of the edge transport and establishes symmetry-derived constraints relevant for spintronic device design.

\indent While this modeling framework captures spin–orbit–induced features relevant for magnetotransport, it is important to note that alternative mechanisms for the topological gap formation have also been proposed. In particular, recent studies suggest that an excitonic condensate may emerge in monolayer 1T$'$-WTe$_2$, driven by strong Coulomb interactions and reduced dielectric screening~\cite{sun2022evidence,kwan2021theory,jia2022evidence,que2024gate,wu2024quasiparticle}. Since the tight-binding model employed here is directly fit to ARPES spectra~\cite{lau2019influence}, it does not distinguish between a gap arising from spin–orbit interaction and one originating from excitonic order, at least within a mean-field framework and at low temperatures. This justifies the use of such a model for symmetry-resolved magnetotransport simulations. However, the coupling between excitonic correlations and spin-polarized edge transport remains an open question for future investigation.


\begin{figure*}
	\includegraphics[width=\linewidth]{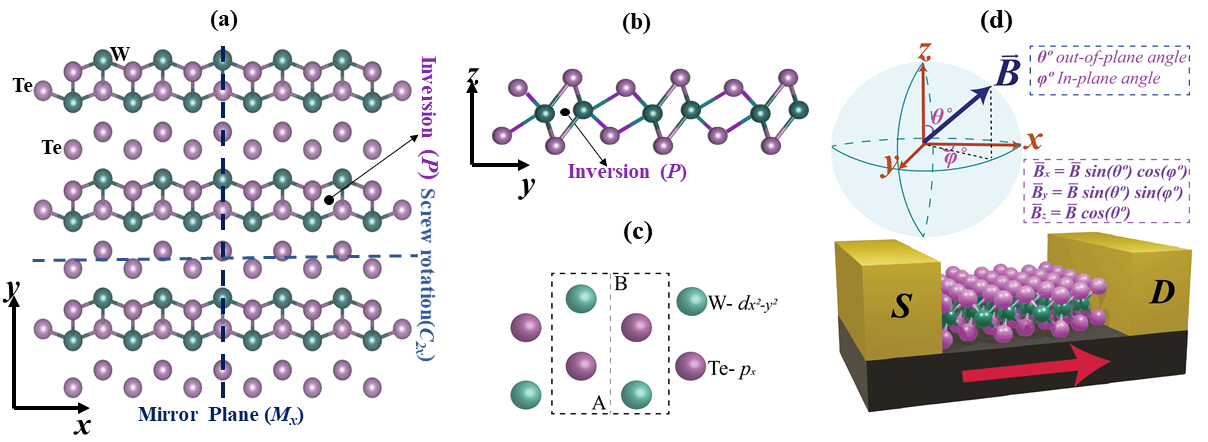}
    \caption{Schematics of the unit cell, lattice structure and magnetotransport setup. (a) The unit cell of WTe\textsubscript{2} consists of four atoms—two W and two Te—where the dominant contributing orbitals are $d_{x^2-y^2}$ (W) and $p_x$ (Te). 
    (b) Side view illustrating the distorted lattice structure. (c) Top view depicting key symmetries, including skew two-fold rotation ($\bar C_{2x}$) along the $x$-axis,  glide mirror symmetry ($ \bar M_x$) in the $y$-$z$ plane, and an inversion center ($P$) arising from the combination of these symmetries.(d) Device schematic for magnetotransport simulations in a two-terminal system. The source (S) and Drain (D) electrodes are aligned either along the W atomic chain ($\hat{x}$-axis) or perpendicular to it ($\hat{y}$-axis). An external magnetic field $\mathbf{B}$ is applied at a polar angle $\theta$ and an azimuthal angle $\varphi$ with respect to the $\hat{x}$-axis, resulting in finite components along all spatial directions.}
    \label{fig:Fig_1}
\end{figure*} 

\indent With the aforesaid, our simulations uncover a pronounced spin anisotropy between transport directions aligned parallel and perpendicular to the W atomic chains. In nanoribbons with edges along the $\hat{y}$-axis (perpendicular to the chains), we find significant spin splitting and an out-of-plane, $\hat{y}$–$\hat{z}$ canted spin texture, consistent with experimental observations~\cite{zhao2021determination,tan2021spin}. In contrast, $\hat{x}$-oriented nanoribbons (along the W atom chains) show negligible spin splitting even under applied magnetic fields. This contrast originates from glide mirror and screw rotation symmetries, which protect band degeneracies along the $\Gamma$–$X$ direction~\cite{symm_theory1,symmetry_transport_exp,lau2022nonreciprocal}. These degeneracies are preserved in the presence of SOC and broken time-reversal symmetry, leading to robust spin-degenerate edge states and pronounced magnetotransport anisotropy.

\indent Our findings establish the symmetry-enforced origin of spin anisotropy in monolayer 1T$'$-WTe$_2$ and provide direct transport-level validation of nonsymmorphic protection. The remainder of this paper is organized as follows: Section~\ref{Sec2} introduces the theoretical model and simulation methodology, Section~\ref{Sec3} presents the magnetotransport results and symmetry analysis, and Section~\ref{Sec4} concludes with implications for symmetry-guided spintronic device design using low-symmetry two-dimensional materials.

\section{Methodology}\label{Sec2}

\indent Monolayer 1T$^\prime$-WTe\textsubscript{2} is a layered material composed of a three-atomic-layer structure, where a tungsten (W) atomic plane is intercalated between two tellurium (Te) planes. The orthorhombic lattice has in-plane lattice constants $a = 3.477$\,\text{\AA} (parallel to W atomic chains) and $b = 6.249$\,\text{\AA} (perpendicular to the chains). The crystal belongs to the nonsymmorphic space group $P2_1/m$, which combines point group operations with fractional translations. Specifically, the symmetry operations include:  
(i) twofold screw rotation $\bar{C}{2x} = { C{2x} \mid a/2,\hat{x} }$;
(ii) glide mirror symmetry $\bar{M}_x = { M_x \mid a/2,\hat{x} }$;
(iii) inversion symmetry $\mathcal{P}$ centered between the W and Te layers (see Fig.~\ref{fig:Fig_1}(a)).
\indent These symmetries constrain the band structure and enforce symmetry-protected degeneracies along high-symmetry directions in the Brillouin zone, notably along the $\Gamma$--$X$ path. They also govern spin selection rules, texture orientation, and anisotropic magnetotransport behavior under spin-orbit coupling (SOC) and time-reversal symmetry breaking.

The low-energy electronic structure is dominated by W $3d_{x^2 - y^2}$ and Te $5p_x$ orbitals, as established by density functional theory (DFT) and angle-resolved photoemission spectroscopy (ARPES). These orbitals form the basis of a minimal tight-binding model consisting of four orbitals per unit cell, as shown in Fig.~\ref{fig:Fig_1}(c), adapted from Ref.~\cite{lau2019influence}, which incorporates the essential topological features and spin-orbit coupling (SOC) effects. The Hamiltonian is given by:

\begin{equation}
\begin{aligned}
H &= \sum_{i,\alpha,s} \epsilon_\alpha\, c_{i\alpha s}^\dagger c_{i\alpha s} 
+ \sum_{\langle i,j \rangle} t^{(1)}_{i\alpha s, j\beta s'}\, e^{i\varphi_{ij}}\, c_{i\alpha s}^\dagger c_{j\beta s'} \\
&\quad + \sum_{\langle\langle i,j \rangle\rangle} t^{(2)}_{i\alpha s, j\beta s'}\, e^{i\varphi_{ij}}\, c_{i\alpha s}^\dagger c_{j\beta s'},
\end{aligned}
\end{equation}
where $c_{i\alpha s}^\dagger$ creates an electron with spin $s$ in orbital $\alpha$ at site $i$. The hopping terms $t^{(1)}$ and $t^{(2)}$ include both spin-conserving and spin-flip processes arising from intrinsic SOC. Model parameters, including onsite energies $\epsilon_\alpha$ and hopping amplitudes, are obtained from maximally localized Wannier functions constructed from DFT. The gap between the valence and conduction bands has been experimentally confirmed by scanning tunneling spectroscopy (STS) to be 56\,meV~\cite{symmetry_transport_exp}.

In order to evaluate the directional anisotropy of edge states along different crystallographic orientations, we perform magnetotransport simulations under an external magnetic field. The magnetic field is incorporated via two contributions: (i) a Zeeman interaction acting on the spin degree of freedom, and (ii) a coupling between the external magnetic field and the orbital motion of electrons, introduced via the Peierls substitution, which modifies the hopping terms with a phase factor derived from the magnetic vector potential.

The Zeeman interaction is modeled as
\begin{equation}
H_Z = -\frac{g \mu_B}{2} {\sigma} \cdot {B},
\end{equation}
where \( \mu_B = 5.78 \times 10^{-5} \, \text{eV/T} \) is the Bohr magneton, \( \boldsymbol{\sigma} \) is the vector of Pauli matrices, and \( g = 4.7 \) is the effective $g$-factor based on experimental estimates~\cite{shi2019imaging}. The magnetic field is parameterized in spherical coordinates as
\begin{equation}
B_x = B \sin\theta \cos\varphi, \quad 
B_y = B \sin\theta \sin\varphi, \quad 
B_z = B \cos\theta,
\end{equation}
where \( \theta \) is the polar angle (measured from the out-of-plane \(\hat{z} \)-axis) and \( \varphi \) is the azimuthal angle (measured from the \( \hat{x} \)-axis).

For numerical reference, the Zeeman energy splitting \( E_Z = g \mu_B B \) leads to
\begin{equation}
E_Z \, [\text{meV}] \approx 0.27 \times B \, [\text{T}],
\end{equation}
for \( g = 4.7 \). However, following Ref.~\cite{lau2019influence}, we adopt the empirical fit
\begin{equation}
B \, [\text{T}] \approx 4.7 \times E_Z \, [\text{meV}],
\end{equation}
which effectively captures edge-state behavior in monolayer WTe\textsubscript{2}. This approximation holds for moderate field strengths with out-of-plane orientation. At higher fields and tilted field directions, the anisotropic and tensorial nature of the $g$-factor may become relevant.

The applied magnetic field also influences the hopping terms which are included via the Peierls' substitution:
\begin{equation}
t_{ij} \rightarrow t_{ij} \, e^{i \varphi_{ij}}, \quad \varphi_{ij} = \frac{e}{\hbar} \int_{{r}_i}^{{r}_j} {A} \cdot d{l}.
\end{equation}
The magnetic vector potential is defined in the symmetric gauge:
\begin{equation}
{A} = \frac{1}{2} {B} \times {r}.
\end{equation}
This yields the following components:
\begin{align}
A_x &= -\frac{B \cos\theta}{2} \hat{y} + \frac{B \sin\theta \cos\varphi}{2} \hat{z}, \\
A_y &= \frac{B \sin\theta \cos\varphi}{2} \hat{x} - \frac{B \sin\theta \sin\varphi}{2} \hat{z}.
\end{align}
These expressions are used to compute $\varphi_{ij}$ for each hopping path in the tight-binding model.

\subsection*{Scattering Geometry and Transport Calculation}

Magnetotransport simulations are performed in a two-terminal nanoribbon geometry, as illustrated in Fig.~\ref{fig:Fig_1}(d). The setup consists of a central scattering region of size $200 \times 150$ unit cells, with open boundary conditions along the transverse ($\hat{y}$) direction and either open or periodic boundaries along the transport ($\hat{x}$) direction.

We compute the electronic transmission using the scattering matrix formalism, implemented via the \texttt{KWANT} package~\cite{groth2014kwant}, following the approach adopted in our previous studies\cite{tapar2023effectuating,tapar2025inclined}. The total transmission probability at a given energy $E$ is given by: 
\begin{equation}
T(E) = \sum_{n \in L, m \in R} |S_{mn}(E)|^2,
\end{equation}
where $S_{mn}$ is the transmission amplitude, computed from the S-matrix formalism from mode $n$ in the left lead to mode $m$ in the right lead. The Fermi energy is set to $E_F = -0.01$\,eV. Mode- and energy-resolved current densities are also computed to distinguish edge-localized contributions from bulk transport under various magnetic field orientations.

\subsection*{Nonsymmorphic Symmetry Constraints}

Monolayer 1T$^\prime$-WTe\textsubscript{2} belongs to the nonsymmorphic space group $P2_1/m$, whose little group along the $\Gamma$–$X$ direction includes the following operations:
\begin{itemize}
    \item \textit{Glide mirror} $\bar{M}_x = \{ M_x \mid a/2 \}$: a mirror reflection across the $x = 0$ plane followed by a half-lattice translation along the $x$-axis.
    \item \textit{Screw rotation} $\bar{C}_{2x} = \{ C_{2x} \mid a/2 \}$: a $180^\circ$ rotation about the $x$-axis followed by the same half-lattice translation.
\end{itemize}

These operations satisfy the algebraic relations:
\begin{equation}
\bar{M}_x^2 = \bar{C}_{2x}^2 = -e^{-ik_x a}, \qquad \bar{M}_x \bar{C}_{2x} = -\bar{C}_{2x} \bar{M}_x.
\end{equation}

As a result:
\begin{enumerate}
    \item Any Bloch state $|\psi\rangle$ that is an eigenstate of $\bar{C}_{2x}$ with eigenvalue $\lambda$ has a degenerate partner $\bar{M}_x |\psi\rangle$ with eigenvalue $-\lambda$, enforcing twofold degeneracy along the $\Gamma$–$X$ direction even in the presence of broken time-reversal symmetry.
    \item Along $\Gamma$–$Y$, the little group does not contain these nonsymmorphic symmetries. Consequently, SOC and magnetic fields can lift degeneracies and induce spin-polarized states.
\end{enumerate}

This anisotropy, rooted in symmetry-protected or lifted band degeneracies, motivates a closer examination of edge-dependent spin transport. In the following sections, we analyze the band structure and perform magnetotransport simulations to confirm the spin splitting and conductance vary with crystal orientation and magnetic field direction.

\section{Results and Discussion} \label{Sec3}
\begin{figure*}
    \includegraphics[width=\textwidth]{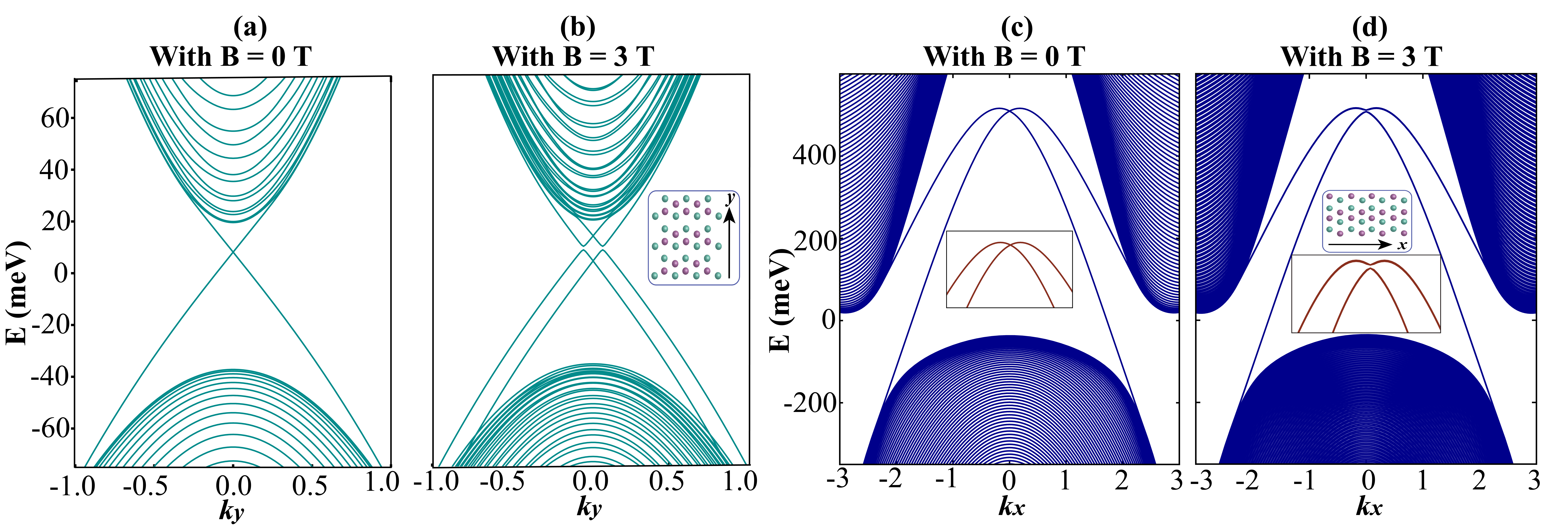}
    \caption{ Band structure of nanoribbons translationally invariant along the $\hat{y}$-axis (a, b) and $\hat{x}$-axis (c, d).
    (a) The band structure of a y-zigzag edge nanoribbon in the absence of a magnetic field shows degenerate edge states with clear band crossings.
    (b) Upon applying an out-of-plane magnetic field B= 3T, spin degeneracy is lifted due to Zeeman splitting, resulting in a band gap opening in the edge states and momentum-space separation of spin bands, consistent with spin–orbit coupling effects.
    (c) The band structure of an $\hat{x}$-edge nanoribbon without magnetic field also exhibits degenerate edge states and band crossings within the bulk spectrum.
    (d) Under the same magnetic field B= 3T, only minor changes are observed: a small bandgap emerges and the edge state splitting remains negligible, as highlighted in the inset.}
    \label{fig:Fig_2}
\end{figure*}

This section analyzes the directional dependence of spin transport in monolayer 1T$^\prime$–WTe\textsubscript{2} ribbons under magnetic field modulation. We first examine the band structures of $\hat{x}$- and $\hat{y}$-edge ribbons, focusing on edge-state splitting with and without out-of-plane magnetic fields. Magnetotransport simulations across polar and azimuthal field angles reveal symmetry-driven anisotropies in transmission. Finally, current density maps at different Fermi levels distinguish edge-localized and bulk-mediated transport, establishing the role of SOC, edge orientation, and crystalline symmetry in spin anisotropy.

\subsection*{Band structure analysis under magnetic field}
We first examine the band structures of monolayer 1T$^\prime$–WTe$_2$ nanoribbons with translational symmetry along the $\hat{y}$ and $\hat{x}$ directions. Figures ~\ref{fig:Fig_2}(a–b) display the electronic spectra for a $\hat{y}$-zigzag edge ribbon and Fig.~\ref{fig:Fig_2}(c-d), an $\hat{x}$-oriented, Te-terminated edge ribbon, both without and with an out-of-plane magnetic field ($B = 3$ T).

\begin{figure*}
    \centering
    \includegraphics[width=\textwidth]{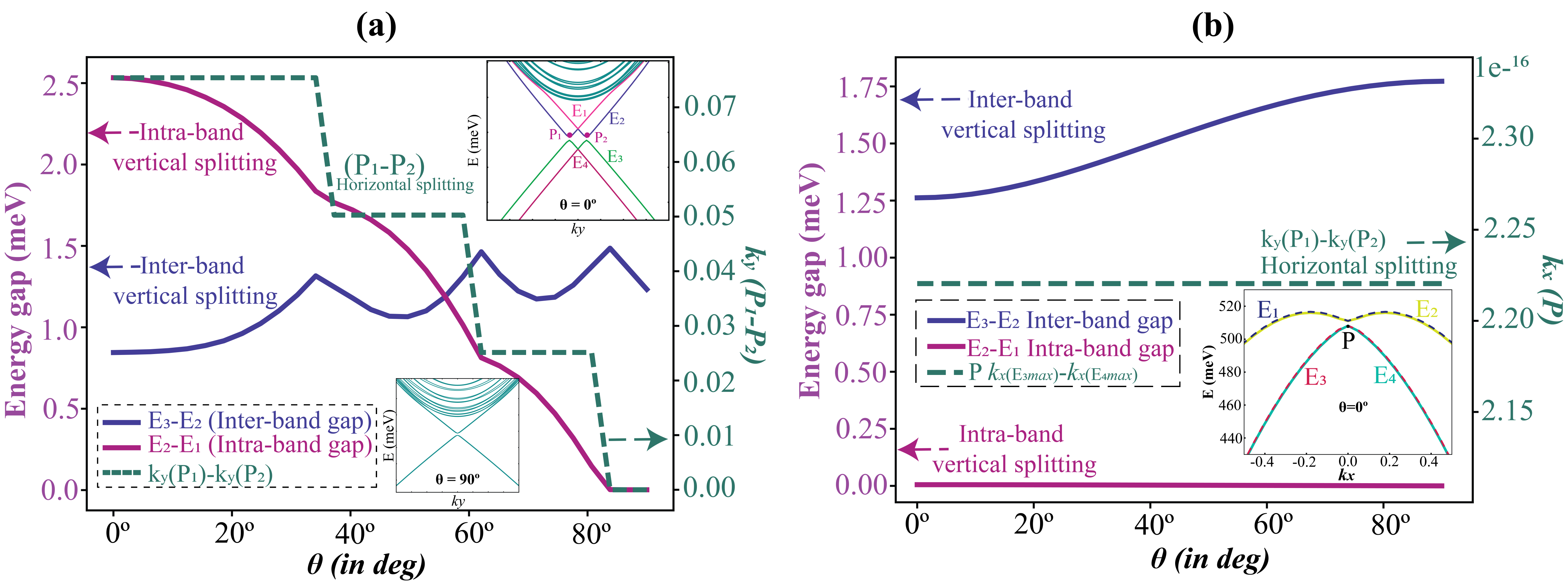}
    \caption {Angle-dependent edge-state splitting in the $\hat{y}$- and $x$-direction nanoribbons.
    Dual-axis plots illustrate the variations in inter- and intra-band splittings in energy (left axis, meV) and momentum (right axis, $k$-units) as a function of the magnetic field polar angle $\theta$ (from $z$-axis to in-plane $\hat{x}$-axis).(a) $\hat{y}$-zigzag ribbon:  At $\theta = 0^\circ$, the inset shows spin-split non-degenerate edge bands. As $\theta$ increases, inter-band energy splitting (dark blue) varies between 0.7–1.5 meV. Intra-band energy splitting (magenta) within the conduction band edge states reduces from $\sim$2.5 meV at $\theta = 0^\circ$ to zero at $\theta = 90^\circ$. A similar decrease is observed for intra-band momentum separation (green). (b) $\hat{x}$- Te terminated ribbon: Intra-band splittings in energy and momentum remain negligible across all $\theta$. A gradual increase in intra-band separation is observed, but without significant spin-resolved splitting. The response remains nearly invariant, indicating minimal sensitivity to magnetic field orientation.}
    \label{fig:Fig_3}
\end{figure*}

Fig.~\ref{fig:Fig_2}(a) shows the band structure of the $\hat{y}$-zigzag edge ribbon in the absence of a magnetic field, where degenerate edge states appear in the mid of the bulk energy gap and exhibit characteristic band crossings. Upon applying an out-of-plane field, as shown in Fig.~\ref{fig:Fig_2}(b), Zeeman interaction and spin-orbit coupling lift the spin degeneracy, resulting in a gap opening in the edge states and a momentum-space separation of spin-split bands. This magnetic-field-induced modification is prominent in this $\hat{y}$-edge configuration.

In contrast, the $\hat{x}$-edge ribbon exhibits degenerate edge states with crossings embedded within the projected bulk spectrum, as shown in Fig.~\ref{fig:Fig_2}(c), with a zoomed-in view of the edge crossing included in the inset. Upon applying the same out-of-plane magnetic field [Fig.~\ref{fig:Fig_2}(d)], only minor modifications are observed: a slight bandgap opens, and spin splitting remains weak, barely lifting the edge-state degeneracy, as highlighted in the inset. This minimal response is consistent with the protection offered by nonsymmorphic symmetries along the $\hat{x}$-direction. 

\indent To examine the angular dependence of edge-state response, we analyze the band structure evolution as a function of the polar angle $\theta$ under a fixed magnetic field of $B = 3$~T. Both Peierls phase and directional Zeeman contributions are consistently included for ribbons with $\hat{x}$- and $\hat{y}$-oriented edges.

\indent As seen earlier in Fig.~\ref{fig:Fig_2}(b), the $\hat{y}$-edge ribbon exhibits notable changes in its edge-state dispersion under an out-of-plane field ($\theta = 0^\circ$). To track this evolution with varying $\theta$, we evaluate three key quantities characterizing the angular response.

First, we examine the inter-band energy splitting, defined as the energy difference between the minimum of the $E_2$ band and the maximum of the $E_3$ band, as shown in the upper inset in Fig.\ref{fig:Fig_3}(a). This splitting, shown by the blue curve, oscillates between 1 and 1.5~meV as $\theta$ varies, indicating a sensitive dependence on the field direction.

Second, the intra-band energy splitting is quantified as the energy difference between the minima of the $E_1$ and $E_2$ bands (or equivalently between the maxima of $E_3$ and $E_4$), and is shown by the magenta curve. This splitting is largest at $\theta = 0^\circ$, where the field is out-of-plane, and gradually reduces to nearly zero at $\theta = 90^\circ$, indicating that spin splitting is strongly suppressed for in-plane magnetic fields.

Finally, we quantify the momentum splitting between spin-split bands by measuring the difference in wavevectors $k_y(P_1)$ and $k_y(P_2)$. Half of this value corresponds to the offset between the band minima of $E_1$ and $E_2$ located symmetrically on either side of the Brillouin zone center. This momentum separation also reaches its maximum at $\theta = 0^\circ$ and diminishes toward $\theta = 90^\circ$, following the same trend as the energy splittings.\indent These results establish that edge-state spin splitting is maximized for $\theta = 0^\circ$ (out-of-plane field) and reduces to a band gap opening for $\theta = 90^\circ$ (in-plane field). Note: Ref.~\cite{lau2019influence} also reports band structure under out-of-plane magnetic fields using the same tight-binding model, but without including the Peierls phase, resulting in qualitatively different edge-state features.

Next, we examine the angular response of the $\hat{x}$-edge (Te-terminated) ribbon using the same analysis framework applied to the $\hat{y}$-edge case. The band structure in Fig.~\ref{fig:Fig_2}(d) shows that applying an out-of-plane magnetic field induces a small energy gap at the edge state crossing. To quantify its evolution with respect to the polar angle $\theta$, Fig.~\ref{fig:Fig_3}(b) presents the extracted inter- and intra-band splittings. The inter-band energy gap, defined as the energy difference between the minimum of the $E_2$ band and the maximum of the $E_3$ band, is shown in blue and increases gradually from approximately 1.25 to 1.75~meV as $\theta$ increases.

In contrast to the $y$-edge ribbon, the intra-band splittings are weak across all $\theta$. The energy separation between the minima of the $E_1$ and $E_2$ bands near $k_y = 0$ (magenta curve) and the momentum displacement of the band extremum $P$ from $k_y = 0$ (green curve) both remain negligible and show no evidence of strong spin band splitting. This limited response to the magnetic field orientation confirms the symmetry-protected nature of edge states along the $x$-direction.

\subsection*{Magneto-transport Simulations}

\begin{figure*}[!t]
    \centering
    \includegraphics[width=\linewidth]{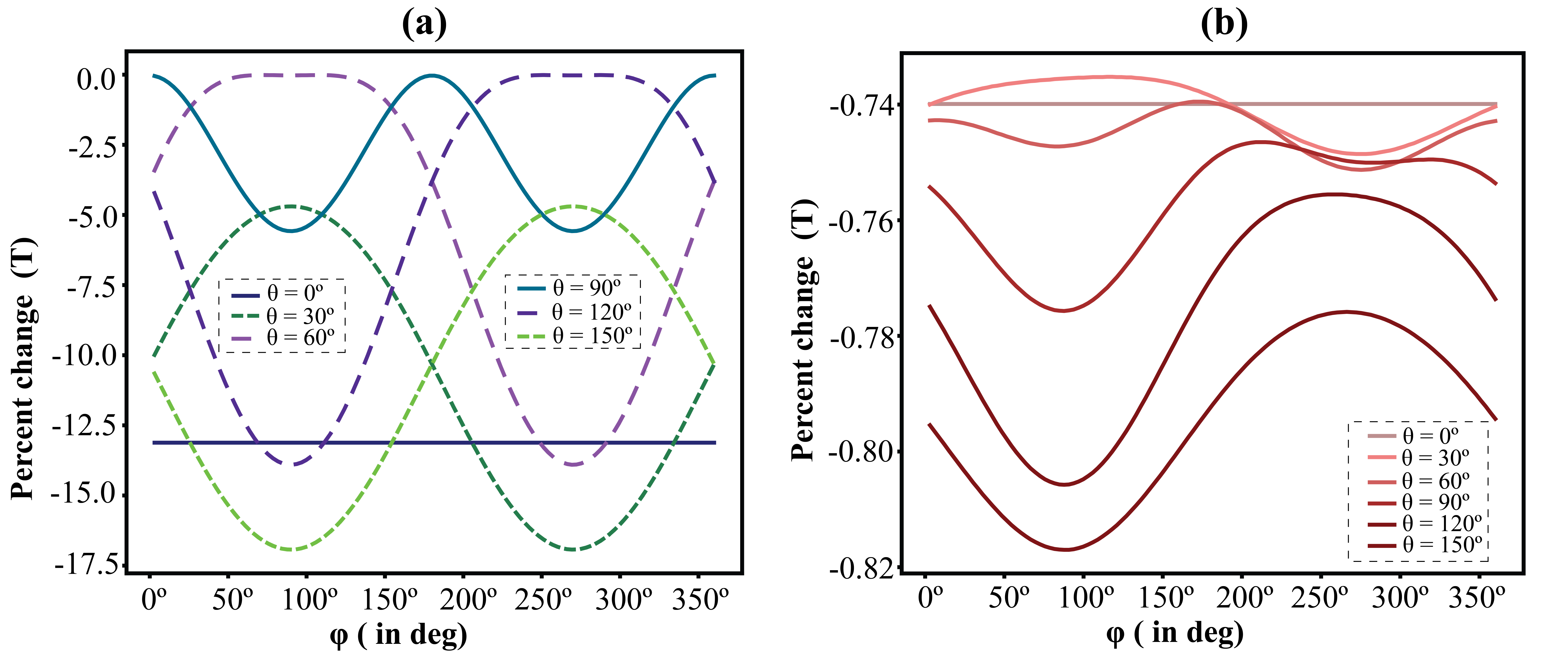}
    \caption{  Magnetotransport simulations for complete in-plane angle (\( \varphi \)) variation at different out-of-plane angles (\( \theta \)) for y- and x-edge ribbons. 
    (a) For the $\hat{y}$-edge ribbon, characteristic antisymmetric transmission behavior is observed for angle pairs such as \( \theta = 30^\circ \) and \( 150^\circ \), as well as \( \theta = 60^\circ \) and \( 120^\circ \), with respect to \( \varphi = 180^\circ \). This confirms the role of mirror symmetry (across the $\hat{y}-\hat{z}$ plane) and spin-momentum locking. The percent change in transmission remains significant, comparable to Fig.~4(a). At \( \theta = 90^\circ \), symmetric minima appear at \( \varphi = 90^\circ \) and \( 270^\circ \), corresponding to magnetic field orientation orthogonal to current. 
    (b) For the $\hat{x}$-edge ribbon, transmission variation remains below 1\%, showing weak magnetotransport response across \( \varphi \), consistent with symmetry-protected band degeneracy and the absence of edge-state modulation in the x-direction.}
    \label{fig:Fig_4}
\end{figure*}

\begin{figure*}[!t]
    \centering
    \includegraphics[width=\linewidth]{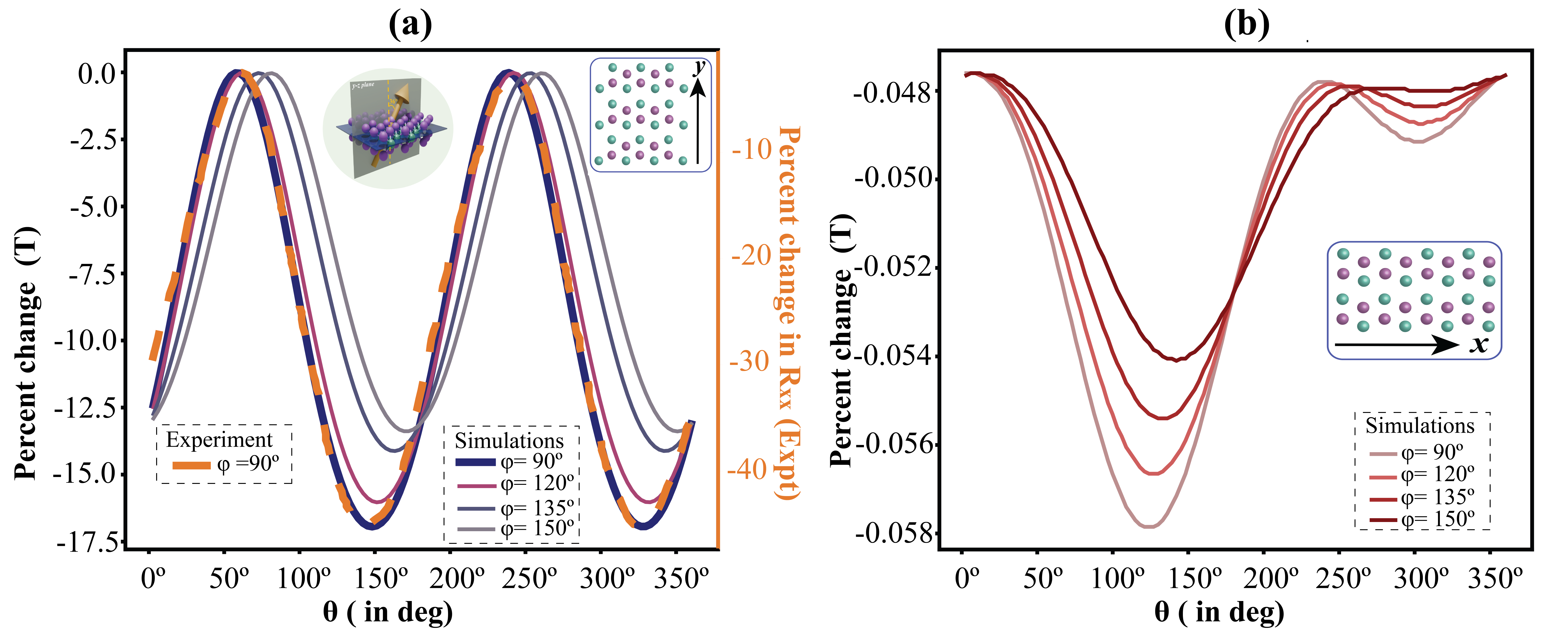}
    \caption{Magnetotransport response as a function of polar angle $\theta$ for different azimuthal angles $\varphi$ in $\hat{y}$- and $\hat{x}$-edge ribbons.
    (a) Dual-axis plot for the $\hat{y}$-edge ribbon: simulated transmission change (left axis) and experimental resistance variation (right axis, ref.\cite{tan2021spin}, Fig. 2(d) show excellent agreement, particularly for $\varphi = 90^\circ$, validating the spin quantization direction at $\theta = 146^\circ$. Other $\varphi$-curves also reproduce experimental trends at $B = 3$~T (not shown here). Simulation range spans 0–18\% change, while experimental variations reach up to 50\%.(b) For the $\hat{x}$-edge ribbon, simulated transmission changes remain below 0.5\%, consistent with the suppressed angular sensitivity due to nonsymmorphic symmetry–protected band degeneracy observed in band structure analysis.}
    \label{fig:Fig_5}
\end{figure*}

We analyze angular magnetotransport by simulating electron transmission in $\hat{x}$- and $\hat{y}$-edge ribbons under an external magnetic field. Transmission is computed as a function of the azimuthal angle $\varphi$ at fixed polar angles $\theta$ (Fig.~\ref{fig:Fig_4}), followed by polar angle variation at selected $\varphi$ values (Fig.~\ref{fig:Fig_5}). The simulations are performed at a Fermi energy of $E_f = -0.01$~eV, where edge-state conduction dominates, and under a magnetic field of $B = 3$~T. Transmission is plotted as a percentage change relative to the maximum possible value of 2, which corresponds to full transmission through both of the edge modes. For the $\hat{y}$-edge ribbon, where transport occurs along the $\hat{y}$-axis, the response is inherently phase-shifted by $90^{\circ}$ relative to the $\hat{x}$-axis reference.
Figure~\ref{fig:Fig_4}(a) shows the simulated transmission for the $\hat{y}$-zigzag edge ribbon as a function of $\varphi$ for different values of $\theta$. At $\theta = 90^{\circ}$, prominent dips occur at $\varphi = 90^{\circ}$ and $270^{\circ}$, aligning with experimental observations in Ref.~\cite{tan2021spin} Fig. 2(b), and consistent with a field applied orthogonal to the transport direction. At other angles, such as $\theta = 30^{\circ}$ and $150^{\circ}$, or $\theta = 60^{\circ}$ and $120^{\circ}$, the transmission curves exhibit antisymmetric behavior across $\varphi = 180^{\circ}$. This antisymmetry originates from the mirror relation between these angle pairs with respect to the $\hat{x}$-axis.

In 1T$^\prime$-WTe\textsubscript{2}, the glide mirror symmetry $\bar{M}_x$, defined as reflection through the $y$-$z$ plane, plays a central role in shaping the transport behavior. Under this operation, the polar vector the $k_x$ changes sign, while the axial spin component $s_x$ remains unchanged. This inversion modifies the spin-momentum locking for mirror-symmetric magnetic field orientations, such as angle pairs $\theta = 30^\circ$ and $150^\circ$, which are symmetric across $\varphi = 180^\circ$. As a result, the transmission response becomes antisymmetric with respect to azimuthal angle. In simulations, the observed transmission variation reaches up to 17\% of the maximum value, highlighting the strong symmetry-governed angular dependence.

In contrast, Fig.~\ref{fig:Fig_4}(b) presents the transmission for the $\hat{x}$-edge (Te-terminated) ribbon under identical simulation parameters. Here, the overall change remains below 1\% across all polar angles, and minor features appear at $\varphi = 90^{\circ}$ for $\theta = 90^{\circ}$ and $270^{\circ}$. This negligible variation is consistent with the suppressed spin splitting observed in the $\hat{x}$-edge band structure and reflects the symmetry-protected nature of edge states along this orientation.

Next, we consider the variation of polar angle $\theta$ for different fixed values of azimuthal angle $\varphi$. In Fig.~\ref{fig:Fig_5}(a), we present a dual-$\hat{y}$-axis plot for the $\hat{y}$-edge ribbon, where the left axis shows the simulated transmission (percent change), and the right axis displays experimental resistance data from Ref.~\cite{tan2021spin}, extracted from Fig.~2(d) at $\varphi = 90^\circ$. Notably, the simulated transmission curve at $\varphi = 90^\circ$ exhibits excellent agreement with the experimental $R_{xx}$ response.

\indent Both the experimental $R_{xx}(\theta)$ and the simulated $T(\theta)$ exhibit sinusoidal angular dependence, with maxima and minima separated by 90$^\circ$ within each curve. Because of the inverse relationship $R \sim 1/T$, the dips in $R_{xx}$ correspond to peaks in $T$. Moreover, since transport is simulated along the $\hat{y}$-axis and measured along the $\hat{x}$-axis in the experiment, the 90$^\circ$ geometric offset adds to the phase relation. Taken together, the total phase shift between the two curves is effectively 180$^\circ$ (or equivalently 0$^\circ$, due to periodicity), allowing for a direct and valid benchmarking of simulation against experiment.

The simulations also reproduce the spin quantization axis orientation at $\theta = 146^\circ$, consistent with Refs.~\cite{tan2021spin, zhao2021determination}. The computed transmission variation spans 0-17\%, while the corresponding experimental resistance varies from 0 to 50\%.

For the $x$-edge ribbon, magnetotransport simulations were also carried out over the same $\theta$ variation at different $\varphi$ values as shown in Fig.\ref{fig:Fig_5}(b). No experimental reference is available for this configuration, as most studies focus on transport perpendicular to the W-atom chains (along the $\hat{y}$-axis). The observed change in transmission relative to the maximum value (approximately 2) remains below 0.5\%, indicating a significantly weaker angular response compared to the $\hat{y}$-edge case.

\begin{figure*}[!t]
    \centering
    \includegraphics[width=\linewidth]{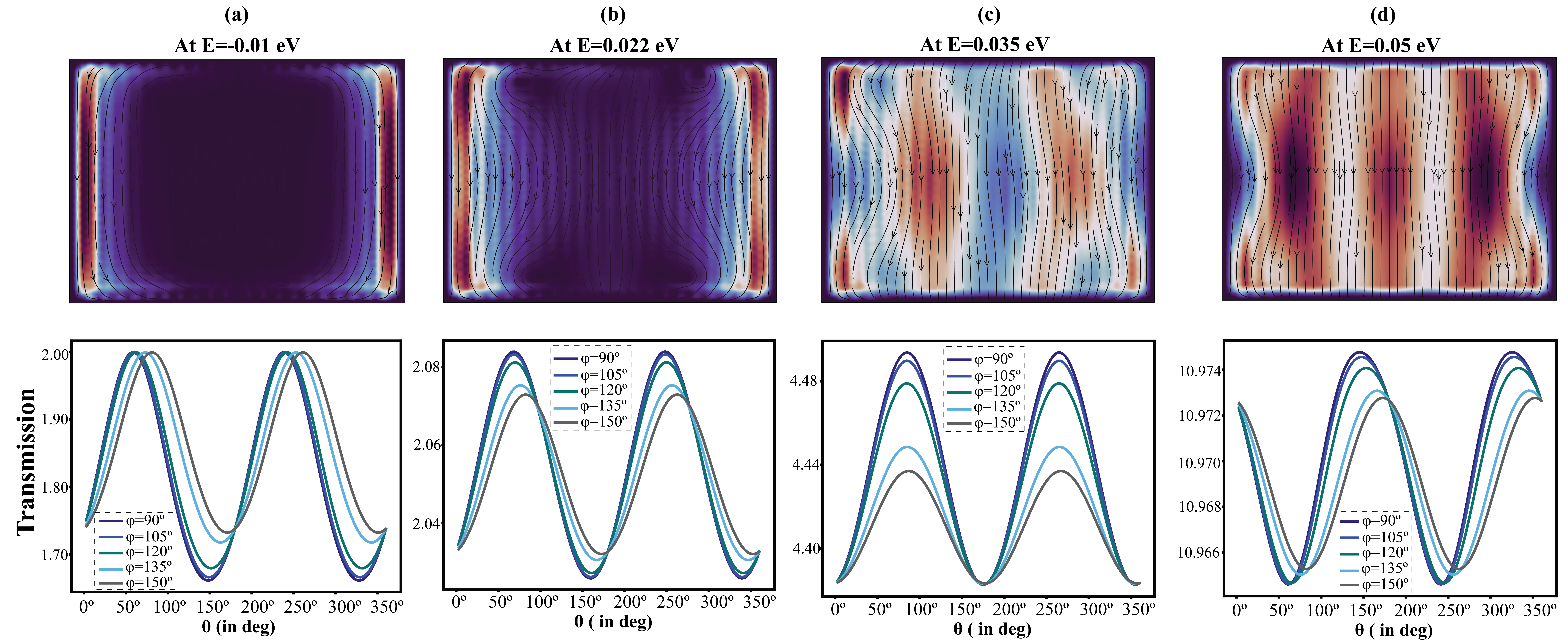}
    \caption{Current density and magnetotransport for the $\hat{y}$-edge zigzag ribbon at different Fermi energies. The device dimensions for the current density plots are $180 \times 200$ nm. (a) At $E = -0.01$ eV, transport is dominated by edge states, with current confined to the edges and negligible bulk contribution. Magnetotransport exhibits strong angular dependence, showing a pronounced minimum near $\theta = 146^\circ$, indicating a preferred spin orientation. (b) At $E = 0.022$ eV, both edge and a few bulk modes contribute. The current partially penetrates the bulk, and the angular variation in magnetotransport is reduced. (c) At $E = 0.035$ eV and $E = 0.05$ eV, transport is bulk-dominated with minimal edge current. The magnetotransport response becomes nearly isotropic, confirming that the model selectively probes edge-state-driven transport at low energies.}
    \label{fig:Fig_6}
\end{figure*}

\begin{figure*}[!t]
    \centering
    \includegraphics[width=\linewidth]{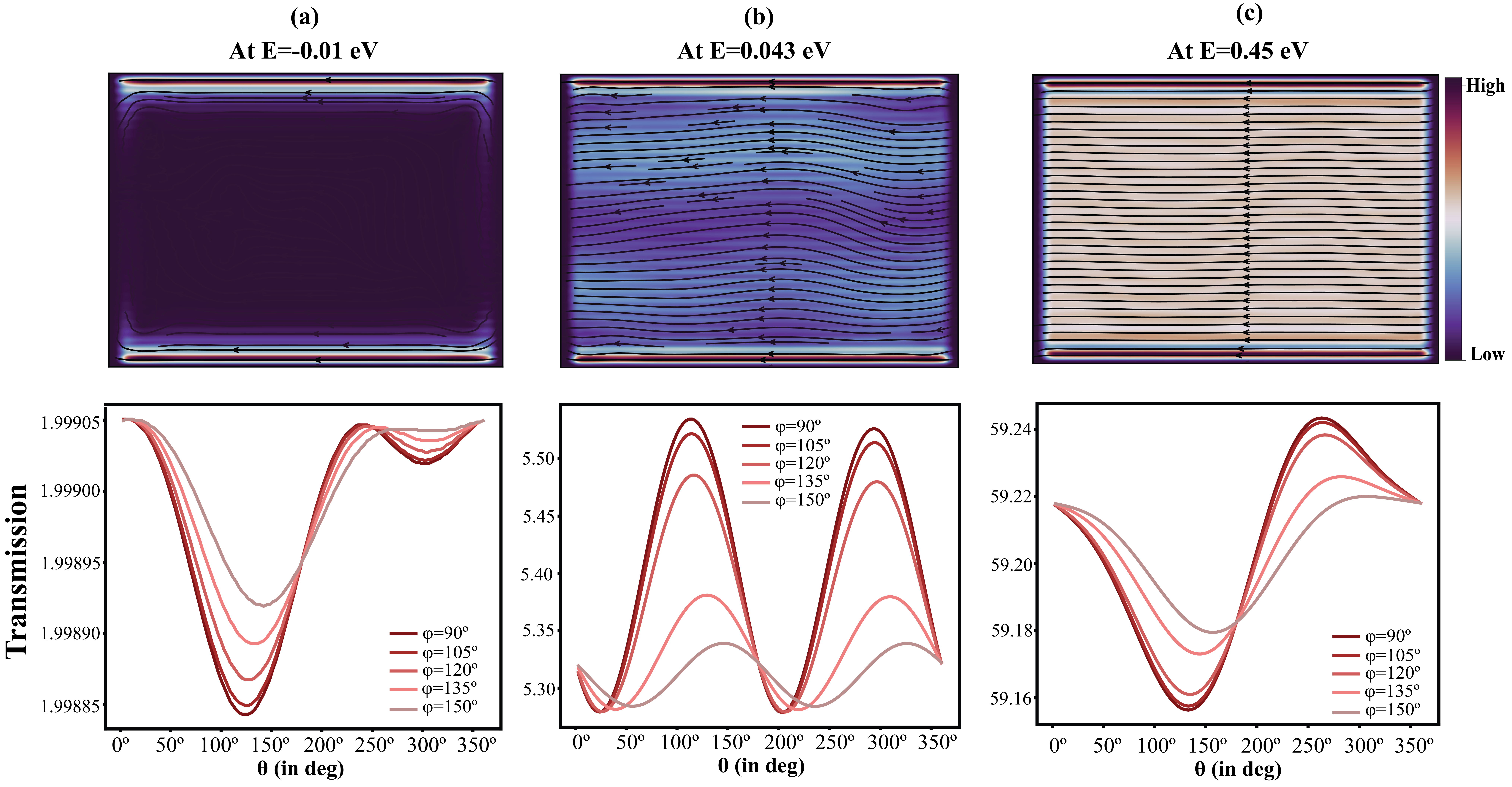}
    \caption{Current density and magnetotransport for the $\hat{x}$-edge ribbon at different Fermi energies (device size: $180 \times 200$ nm). (a) At $E = 0.01$ eV, transport is edge-state–dominated, with current confined to the ribbon boundaries. Magnetotransport shows negligible angular dependence, consistent with symmetry-protected spin degeneracy. (b, c) At $E = 0.043$ eV and $E = 0.45$ eV, bulk modes dominate, as seen from current spreading into the interior. A slight increase in angular variation appears, similar to the bulk-dominated regime in the $\hat{y}$-edge ribbon. These results support the model's ability to isolate edge-state magnetotransport signatures.}
    \label{fig:Fig_7}
\end{figure*}

Based on magnetotransport simulations focused on the response of edge states to magnetic field variation in both $\hat{y}$- and $\hat{x}$-edge ribbons, the out-of-plane spin orientation axis is confirmed from the $y$-edge transport results. Additionally, the symmetry-enforced spin degeneracy—arising from nonsymmorphic symmetries, is validated through both the band structure and angular magnetotransport behavior.

Next, we examine the magnetotransport characteristics as a function of Fermi energy to analyze the transition from edge-dominated to bulk-dominated transport. This allows us to differentiate the transport response of edge states and bulk states and to compare their behavior in both $\hat{y}$- and $\hat{x}$-edge configurations.

Figure \ref{fig:Fig_6} presents spatial current‐density maps and the corresponding magnetotransport response for the $\hat{y}$‐zigzag edge ribbon, evaluated at several Fermi energies as a function of the polar angle $\theta$ (for selected azimuthal angles $\varphi$).

At $E_{\!f} = -0.01$~eV, the current density is confined to the ribbon edges, indicating edge-state-dominated transport. The associated transmission exhibits strong angular dependence, with percent changes on the order of $10^{-1}$. Increasing the Fermi level to $E_{\!f} = 0.022$~eV introduces partial bulk conduction, and the corresponding transmission variation reduces to the order of $10^{-2}$, as shown in Fig.~\ref{fig:Fig_6}(b). At higher energies, $E_{\!f} = 0.035$~eV [Fig.~\ref{fig:Fig_6}(c)] and $E_{\!f} = 0.05$~eV [Fig.~\ref{fig:Fig_6}(d)], bulk states dominate the transport. The current density becomes nearly uniform across the ribbon width, and the angular variation in transmission further diminishes to the $10^{-2}$ and $10^{-3}$ ranges, respectively. This systematic reduction in magneto-angular response highlights the crucial role of edge states in mediating spin–orbit-coupled, field-sensitive transport in 1T$^\prime$–WTe\textsubscript{2}.

Now, considering the $x$-edge Te-terminated ribbon, we begin by analyzing transport at $E_{\!f} = -0.01$~eV. As shown in the current density plot [Fig.~\ref{fig:Fig_7}(a)], the transport is dominated by edge states. However, as discussed earlier in Fig.~\ref{fig:Fig_5}(b), the corresponding change in transmission remains minimal—below 1\%—and falls in the range of $10^{-5}$. As the Fermi energy increases and bulk states begin to contribute, the current density transitions toward a more uniform distribution, as illustrated in Fig.~\ref{fig:Fig_7}(b) and (c) for $E_{\!f} = 0.043$~eV and $E_{\!f} = 0.45$~eV, respectively. At these energies, the transmission variation due to magnetic field orientation rises to the order of $10^{-2}$, consistent with the behavior observed in the $\hat{y}$-edge ribbon under bulk-dominated conditions.

As observed for both $\hat{y}$- and $\hat{x}$-edge ribbons, bulk states exhibit negligible sensitivity to variations in the applied magnetic field. This behavior aligns with experimental findings reported in Ref.~\cite{zhao2021determination}, which show that at elevated temperatures (e.g., 140~K), bulk conduction becomes dominant and remains largely unaffected by magnetic field modulation.

\section{Conclusion} \label{Sec4}
This work investigates the directional anisotropy in magnetotransport of monolayer 1T$^\prime$-WTe\textsubscript{2} by analyzing the angular dependence of electronic transmission in $\hat{y}$- and $\hat{x}$-ribbon geometries under an external magnetic field. The $y$-ribbon exhibits strong angular modulation of conductance, which correlates with magnetic-field-induced band splitting of edge states in both energy and momentum space. In contrast, the $\hat{x}$-ribbon shows negligible angular response, consistent with the preservation of band degeneracy enforced by the material’s non-symmorphic symmetry. Glide mirror and screw rotation operations protect the electronic structure along the $\hat{x}$-direction from field-induced splitting, even in the absence of time-reversal symmetry. Energy-resolved analysis confirms that edge contributions dominate this anisotropic behavior, while the bulk remains largely insensitive to field orientation. These findings provide a transport-based signature of non-symmorphic-symmetry-protected degeneracies in 1T$^\prime$-WTe\textsubscript{2} and offer design guidelines for engineering directional control in quantum devices based on lattice symmetries.
\begin{acknowledgments}
We thank Dr.\ Anirban Das for his valuable insights and helpful discussions, particularly during the initial setup of the theoretical and simulation framework used in this work.
The author BM acknowledges funding from the Department of Science and Technology (DST), Government of India, under the National Quantum Mission, {through Grant no. DST/QTC/NQM/QMD/2024/4}, the Institute of Eminence Grant from IIT Bombay, {through Grant no. RD/0523-IOE00I0-139} and the Inani Chair Professorship fund {, through Grant No. DO/2024-INAN/001-001}. The authors BM and ST acknowledge funding from the Dhananjay Joshi Endowment award from IIT Bombay{, through Grant No: DO/2023-DJEF002}. The author BW acknowledges the support of the National Research Foundation (NRF) Singapore, under the Competitive Research Program “Towards On-Chip Topological Quantum Devices” (NRF-CRP21-2018-0001), with further support from the Singapore Ministry of Education (MOE) Academic Research Fund Tier 3 grant (MOE-MOET32023-0003) “Quantum Geometric Advantage” and the Air Force Office of Scientific Research under award number FA2386-24-1-4064. Also, acknowledges National Research Foundation, Singapore(Grant No. NRF-F-CRP-2024-0012).\\
\end{acknowledgments}

\textbf{Data availability}
The data that support the findings of this study are available from the corresponding author upon reasonable request.\\
\bibliography{apssamp}

\begin{thebibliography}{10}

\bibitem{muechler2016topological}
Lukas Muechler, Aris Alexandradinata, Titus Neupert, and Roberto Car.
\newblock Topological nonsymmorphic metals from band inversion.
\newblock {\em Physical Review X}, 6(4):041069, 2016.

\bibitem{choe2016understanding}
Duk-Hyun Choe, Ha-Jun Sung, and Kee~Joo Chang.
\newblock Understanding topological phase transition in monolayer transition metal dichalcogenides.
\newblock {\em Physical Review B}, 93(12):125109, 2016.

\bibitem{jia2017direct}
Z.-Y. Jia and et~al.
\newblock Direct visualization of a two-dimensional topological insulator in the single-layer {1T’}-{WTe\textsubscript{2}}.
\newblock {\em Physical Review B}, 96(4):041108, 2017.

\bibitem{qian2014quantum}
Xiaofeng Qian, Junwei Liu, Liang Fu, and Ju~Li.
\newblock Quantum spin hall effect in two-dimensional transition metal dichalcogenides.
\newblock {\em Science}, 346(6215):1344--1347, 2014.

\bibitem{fei2017edge}
Zaiyao Fei, Tauno Palomaki, Sanfeng Wu, Wenjin Zhao, Xinghan Cai, Bosong Sun, Paul Nguyen, Joseph Finney, Xiaodong Xu, and David~H Cobden.
\newblock Edge conduction in monolayer {$WTe_2$}.
\newblock {\em Nature Physics}, 13(7):677--682, 2017.

\bibitem{tang2017quantum}
Shujie Tang, Chaofan Zhang, Dillon Wong, Zahra Pedramrazi, Hsin-Zon Tsai, Chunjing Jia, Brian Moritz, Martin Claassen, Hyejin Ryu, Salman Kahn, et~al.
\newblock Quantum spin hall state in monolayer {1T'}-{WTe\textsubscript{2}}.
\newblock {\em Nature Physics}, 13(7):683--687, 2017.

\bibitem{shi2019imaging}
Yanmeng Shi, Joshua Kahn, Ben Niu, Zaiyao Fei, Bosong Sun, Xinghan Cai, Brian~A Francisco, Di~Wu, Zhi-Xun Shen, Xiaodong Xu, et~al.
\newblock Imaging quantum spin hall edges in monolayer {WTe\textsubscript{2}}.
\newblock {\em Science advances}, 5(2):eaat8799, 2019.

\bibitem{wu2018observation}
Sanfeng Wu, Valla Fatemi, Quinn~D Gibson, Kenji Watanabe, Takashi Taniguchi, Robert~J Cava, and Pablo Jarillo-Herrero.
\newblock Observation of the quantum spin hall effect up to 100 kelvin in a monolayer crystal.
\newblock {\em Science}, 359(6371):76--79, 2018.

\bibitem{lodge2021atomically}
M.~S. Lodge, S.~A. Yang, S.~Mukherjee, and B.~Weber.
\newblock Atomically thin quantum spin hall insulators.
\newblock {\em Advanced Materials}, 33(45):2008029, 2021.

\bibitem{jia2022tuning}
Junxiang Jia, Elizabeth Marcellina, Anirban Das, Michael~S Lodge, BaoKai Wang, Duc-Quan Ho, Riddhi Biswas, Tuan~Anh Pham, Wei Tao, Cheng-Yi Huang, et~al.
\newblock Tuning the many-body interactions in a helical luttinger liquid.
\newblock {\em Nature Communications}, 13(1):6046, 2022.

\bibitem{symmetry_transport_exp}
Valla Fatemi and et~al.
\newblock Electrically tunable low-density superconductivity in a monolayer topological insulator.
\newblock {\em Science}, 362:926--929, 2018.

\bibitem{sajadi2018gate}
Ebrahim Sajadi, Tauno Palomaki, Zaiyao Fei, Wenjin Zhao, Philip Bement, Christian Olsen, Silvia Luescher, Xiaodong Xu, Joshua~A Folk, and David~H Cobden.
\newblock Gate-induced superconductivity in a monolayer topological insulator.
\newblock {\em Science}, 362(6417):922--925, 2018.

\bibitem{xie2020spin}
Ying-Ming Xie, Benjamin~T Zhou, and Kam~Tuen Law.
\newblock Spin-orbit-parity-coupled superconductivity in topological monolayer {WTe\textsubscript{2}}.
\newblock {\em Physical Review Letters}, 125(10):107001, 2020.

\bibitem{song2024unconventional}
Tiancheng Song, Yanyu Jia, Guo Yu, Yue Tang, Pengjie Wang, Ratnadwip Singha, Xin Gui, Ayelet~J Uzan-Narovlansky, Michael Onyszczak, Kenji Watanabe, et~al.
\newblock Unconventional superconducting quantum criticality in monolayer {$WTe_2$}.
\newblock {\em Nature Physics}, 20(2):269--274, 2024.

\bibitem{que2024gate}
Yande Que, Yang-Hao Chan, Junxiang Jia, Anirban Das, Zhengjue Tong, Yu-Tzu Chang, Zhenhao Cui, Amit Kumar, Gagandeep Singh, Shantanu Mukherjee, et~al.
\newblock A gate-tunable ambipolar quantum phase transition in a topological excitonic insulator.
\newblock {\em Advanced Materials}, 36(7):2309356, 2024.

\bibitem{ji2023influence}
Shaozheng Ji, Oscar Gr{\aa}n{\"a}s, Amit~Kumar Prasad, and Jonas Weissenrieder.
\newblock Influence of strain on an ultrafast phase transition.
\newblock {\em Nanoscale}, 15(1):304--312, 2023.

\bibitem{zhao2020strain}
C.~Zhao and et~al.
\newblock Strain tunable semimetal–topological-insulator transition in monolayer {1T’}-{$WTe_2$}.
\newblock {\em Physical Review Letters}, 125(4):046801, 2020.

\bibitem{maximenko2022nanoscale}
Yulia Maximenko, Yueqing Chang, Guannan Chen, Mark~R Hirsbrunner, Waclaw Swiech, Taylor~L Hughes, Lucas~K Wagner, and Vidya Madhavan.
\newblock Nanoscale studies of electric field effects on monolayer {1T'}-{WTe\textsubscript{2}}.
\newblock {\em npj Quantum Materials}, 7(1):29, 2022.

\bibitem{safeer2019sot}
CK~Safeer and et~al.
\newblock Room-temperature spin-orbit torque in van der waals heterostructures.
\newblock {\em Nano Letters}, 19(2):8758--8764, 2019.

\bibitem{macneill2017symmetry}
D~MacNeill and et~al.
\newblock Control of spin–orbit torques through crystal symmetry in {WTe\textsubscript{2}}.
\newblock {\em Nature Physics}, 13(3):300--305, 2017.

\bibitem{macneill2019angular}
David MacNeill and et~al.
\newblock Angular dependence of spin-orbit torques in {WTe\textsubscript{2}}/ferromagnet bilayers.
\newblock {\em Physical Review Letters}, 123:207201, 2019.

\bibitem{wang2019current}
Yujing Wang and et~al.
\newblock Current-induced spin polarization in {WTe\textsubscript{2}} monolayers.
\newblock {\em Nature Communications}, 10:1--6, 2019.

\bibitem{zhang2018electrically}
Yang Zhang, Jeroen Van Den~Brink, Claudia Felser, and Binghai Yan.
\newblock Electrically tuneable nonlinear anomalous hall effect in two-dimensional transition-metal dichalcogenides {WTe\textsubscript{2}} and {MoTe\textsubscript{2}}.
\newblock {\em 2D Materials}, 5(4):044001, 2018.

\bibitem{zhao2020observation}
Bing Zhao, Dmitrii Khokhriakov, Yang Zhang, Huixia Fu, Bogdan Karpiak, Anamul~Md Hoque, Xiaoguang Xu, Yong Jiang, Binghai Yan, and Saroj~P Dash.
\newblock Observation of charge to spin conversion in weyl semimetal {WTe\textsubscript{2}} at room temperature.
\newblock {\em Physical review research}, 2(1):013286, 2020.

\bibitem{liu2023crystallographically}
Tian Liu, Arunesh Roy, Jan Hidding, Homayoun Jafari, Dennis~K De~Wal, Jagoda S{\l}awi{\'n}ska, Marcos~HD Guimar{\~a}es, and Bart~J Van~Wees.
\newblock Crystallographically dependent bilinear magnetoelectric resistance in a thin {WTe\textsubscript{2}} layer.
\newblock {\em Physical Review B}, 108(16):165407, 2023.

\bibitem{spin_texture1}
Y.~Zhang and et~al.
\newblock Observation of a hidden spin-polarized state in monolayer {WTe$_2$}.
\newblock {\em Nano Letters}, 21:1529--1535, 2021.

\bibitem{spin_texture2}
D.~MacNeill and et~al.
\newblock Control of spin-orbit torques through crystal symmetry in {WTe$_2$}/ferromagnet bilayers.
\newblock {\em Nature Physics}, 13:300--305, 2017.

\bibitem{shi2019symmetry}
Li-kun Shi and Justin~CW Song.
\newblock Symmetry, spin-texture, and tunable quantum geometry in a {WTe\textsubscript{2}} monolayer.
\newblock {\em Physical Review B}, 99(3):035403, 2019.

\bibitem{garcia2020canted}
Jose~H Garcia, Marc Vila, Chuang-Han Hsu, Xavier Waintal, Vitor~M Pereira, and Stephan Roche.
\newblock Canted persistent spin texture and quantum spin hall effect in {$WTe_2$}.
\newblock {\em Physical Review Letters}, 125(25):256603, 2020.

\bibitem{arora2020cooperative}
Arpit Arora, Li-kun Shi, and Justin~CW Song.
\newblock Cooperative orbital moments and edge magnetoresistance in monolayer {$WTe_2$}.
\newblock {\em Physical Review B}, 102(16):161402, 2020.

\bibitem{liu2014prl}
Junwei Liu and David Vanderbilt.
\newblock Topological phases of {WTe\textsubscript{2}} monolayers.
\newblock {\em Phys. Rev. Lett.}, 114:136601, 2015.

\bibitem{liu2018prx}
Yuan Liu et~al.
\newblock Symmetry analysis of spin texture in {WTe\textsubscript{2}}.
\newblock {\em Phys. Rev. X}, 8:031070, 2018.

\bibitem{lau2019influence}
Alexander Lau, Rajyavardhan Ray, D{\'a}niel Varjas, and Anton~R Akhmerov.
\newblock Influence of lattice termination on the edge states of the quantum spin hall insulator monolayer {1 T'}-{WTe\textsubscript{2}}.
\newblock {\em Physical Review Materials}, 3(5):054206, 2019.

\bibitem{sun2022evidence}
Bosong Sun, Wenjin Zhao, Tauno Palomaki, Zaiyao Fei, Elliott Runburg, Paul Malinowski, Xiong Huang, John Cenker, Yong-Tao Cui, Jiun-Haw Chu, et~al.
\newblock Evidence for equilibrium exciton condensation in monolayer {$WTe_2$}.
\newblock {\em Nature Physics}, 18(1):94--99, 2022.

\bibitem{kwan2021theory}
Yves~H Kwan, T~Devakul, SL~Sondhi, and SA~Parameswaran.
\newblock Theory of competing excitonic orders in insulating {$WTe_2$} monolayers.
\newblock {\em Physical Review B}, 104(12):125133, 2021.

\bibitem{jia2022evidence}
Yanyu Jia, Pengjie Wang, Cheng-Li Chiu, Zhida Song, Guo Yu, Berthold J{\"a}ck, Shiming Lei, Sebastian Klemenz, F~Alexandre Cevallos, Michael Onyszczak, et~al.
\newblock Evidence for a monolayer excitonic insulator.
\newblock {\em Nature Physics}, 18(1):87--93, 2022.

\bibitem{wu2024quasiparticle}
Jinyuan Wu, Bowen Hou, Wenxin Li, Yu~He, and Diana~Y Qiu.
\newblock Quasiparticle and excitonic properties of monolayer {1T'}-{WTe$_2$} within many-body perturbation theory.
\newblock {\em Physical Review B}, 110(7):075133, 2024.

\bibitem{zhao2021determination}
Wenjin Zhao, Elliott Runburg, Zaiyao Fei, Joshua Mutch, Paul Malinowski, Bosong Sun, Xiong Huang, Dmytro Pesin, Yong-Tao Cui, Xiaodong Xu, et~al.
\newblock Determination of the spin axis in quantum spin hall insulator candidate monolayer {WTe\textsubscript{2}}.
\newblock {\em Physical Review X}, 11(4):041034, 2021.

\bibitem{tan2021spin}
Cheng Tan, Ming-Xun Deng, Guolin Zheng, Feixiang Xiang, Sultan Albarakati, Meri Algarni, Lawrence Farrar, Saleh Alzahrani, James Partridge, Jia~Bao Yi, et~al.
\newblock Spin-momentum locking induced anisotropic magnetoresistance in monolayer {WTe\textsubscript{2}}.
\newblock {\em Nano Letters}, 21(21):9005--9011, 2021.

\bibitem{symm_theory1}
Junwei Liu and David Vanderbilt.
\newblock Topological semimetals from nonsymmorphic symmetries.
\newblock {\em Physical Review B}, 90(15):155316, 2014.

\bibitem{lau2022nonreciprocal}
Audrey Lau and et~al.
\newblock Nonreciprocal charge transport across the quantum spin hall edge.
\newblock {\em Nature Nanotechnology}, 17:1026--1031, 2022.

\bibitem{groth2014kwant}
Christoph~W Groth, Michael Wimmer, Anton~R Akhmerov, and Xavier Waintal.
\newblock Kwant: a software package for quantum transport.
\newblock {\em New Journal of Physics}, 16(6):063065, 2014.

\bibitem{tapar2023effectuating}
Shrushti Tapar and Bhaskaran Muralidharan.
\newblock Effectuating tunable valley selection via multiterminal monolayer graphene devices.
\newblock {\em Physical Review B}, 107(20):205415, 2023.

\bibitem{tapar2025inclined}
Shrushti Tapar and Bhaskaran Muralidharan.
\newblock Inclined junctions in monolayer graphene: a gateway toward tailoring valley polarization of dirac fermions.
\newblock {\em Materials for Quantum Technology}, 5(2):026201, 2025.

\end{thebibliography}
\end{document}